\documentclass[]{aa}  

\usepackage{graphicx}
\usepackage{txfonts}
\usepackage[utf8]{inputenc}
\usepackage{amsfonts}
\usepackage{amsmath}
\usepackage{bm}
\usepackage{xcolor}
\usepackage{color}
\usepackage{amssymb}
\usepackage{natbib}
\usepackage{xspace}
\usepackage{ulem}
\usepackage{orcidlink}

\newcommand{\tbn}{$\theta_{Bn}$ }	
\newcommand{\wind}{\textit{Wind}\xspace}

\begin{document} 

\title{Variability in energetic particle observations at strong interplanetary shocks: Multi-spacecraft observations}
\titlerunning{Energetic particles at IP shocks}

\author{D. Trotta\orcidlink{0000-0002-0608-8897}\inst{1}
\and 
T. S. Horbury\orcidlink{0000-0002-7572-4690}\inst{2}
\and 
J. Giacalone\orcidlink{0000-0002-0850-4233}\inst{3}}

\authorrunning{D. Trotta et al.}
\institute{European Space Agency (ESA), European Space Astronomy Centre (ESAC), Camino Bajo del Castillo s/n, 28692 Villanueva de la Cañada, Madrid, Spain\\ \email{domenico.trotta@esa.int}
\and  
The Blackett Laboratory, Department of Physics, Imperial College London, London SW7 2AZ, UK
\and
Lunar and Planetary Laboratory, University of Arizona, Tucson, USA
}

\date{\today}

\abstract
   {Interplanetary (IP) shock waves offer an unparalleled opportunity to directly study the elusive mechanisms of particle acceleration that are pervasive in our Universe.  Novel spacecraft missions, orbiting poorly-explored regions of the heliosphere, opened a new observational window on particle acceleration at IP shocks that is relevant to space and astrophysical plasmas.}
   {We address shock variability and its effects on the production of accelerated particles at different energies. We leveraged three different missions that directly observed a strong IP shock in a range of separations that cannot be achieved with a single mission. We linked spatial shock irregularities and evolutionary effects to the observed energetic particle responses in the shock passage at the three different heliospheric vantage points.}
   {We exploited direct observations of magnetic field, plasma, and energetic particle fluxes from the \textit{Wind} and ACE missions at 1 AU and from the Solar Orbiter spacecraft. They are well-aligned radially at 0.8 AU. We devised a new technique based on the cross-correlation of energetic particle profiles to quantitatively address the variability in the characteristics of energetic particles at different points in space and time.}
   {We show that ions with different energies respond differently to the shock passage in the range of observer separations 0.02 -- 0.2 AU we explored. The shape and behavior of high-energy ($\gtrapprox$ 0.5 MeV) particle profiles vary between the 0.8 and 1 AU observations, and we suggest that this is caused by shock-evolution, in which high-energy particles are produced less efficiently at 1 AU than at 0.8.  Finally, we show that shock and ambient  spatial irregularities that are observed throughout the event modulate the energetic particle responses at different energies.}
   {}
\keywords{Shock waves -- solar wind  --  Acceleration of particles}

\maketitle

\section{Introduction}
\label{sect:intro}

Shocks are present in many astrophysical systems, where they play a fundamental role in energy conversion~\citep[e.g.,][]{Bykov2019}. Generally speaking, shocks are abrupt transitions between supersonic and subsonic flows, and they convert directed bulk flow energy (upstream) into heat and magnetic energy (downstream). In the collisionless case, a fraction of the available energy can be channeled to the production of energetic particles~\citep[e.g.,][]{Burgess2015}. The mechanisms by which particles are accelerated to high energies are not completely understood, however, and they have been at the center of scientific debate since Fermi's seminal work in the 1950s~\citep{Fermi1949, Fermi1954}. 

Direct observations of shocks and energetic particles are uniquely available in the heliosphere, and they are thus invaluable to understand the governing parameters of efficient particle acceleration~\citep{DesaiGiacalone2016}. They also stimulate cross-disciplinary advances in understanding nonthermal emission from a variety of astrophysical systems, ranging from the solar atmosphere to megaparsec-sized emission at galaxy cluster scales, where only remote observations are possible~\cite{Vrsnak2008,Brunetti2014}. 

The most frequently investigated shock in the heliosphere is the Earth's bow shock, which results from the interaction between the Earth's magnetosphere and the supersonic (super-Alfv\`enic) solar wind~\citep[e.g.,][]{Burgess2012}. Many single- and multi-spacecraft missions have been devoted to investigating the Earth's magnetosphere, and built a comprehensive understanding of how the Earth's bow shock operates from its large-scale behavior to its small-scale kinetic features~\citep{fairfield1976,Lalti2022}. 

Interplanetary (IP) shock waves propagate in interplanetary space due to solar eruptive phenomena such as coronal mass ejections (CMEs) and stream interaction regions (SIRs)~\citep[e.g.,][]{Burlaga1971, Reames1999}. IP shocks offer an invaluable research opportunity that allows us to explore poorly investigated shock parameter regimes that typically are weaker than Earth's bow shock and have larger curvature radii~\citep{Kilpua2015_shocks}. IP shocks are less frequently investigated than the Earth's bow shock because observations are challenging: Their higher speeds constrain the time resolution that is needed to resolve their structure, and fewer multi-spacecraft missions are available to investigate them~\citep{Cohen2019}.

Importantly, by studying how IP shocks propagate in the heliosphere, it is possible to address their evolutionary features and associated particle acceleration~\citep{PerezAlanis2023}, which is not possible for Earth's bow shock. From this point of view, multipoint observations are crucial for building a comprehensive picture of how energetic particle production operates for different evolutionary stages of shocks and to determine the role of spatial irregularities of the shocks in shaping particle acceleration, as shown by observations and simulations~\citep{Lugaz2024,Wijsen2023}.

 \begin{figure*}  
   \includegraphics[width=0.99\textwidth]{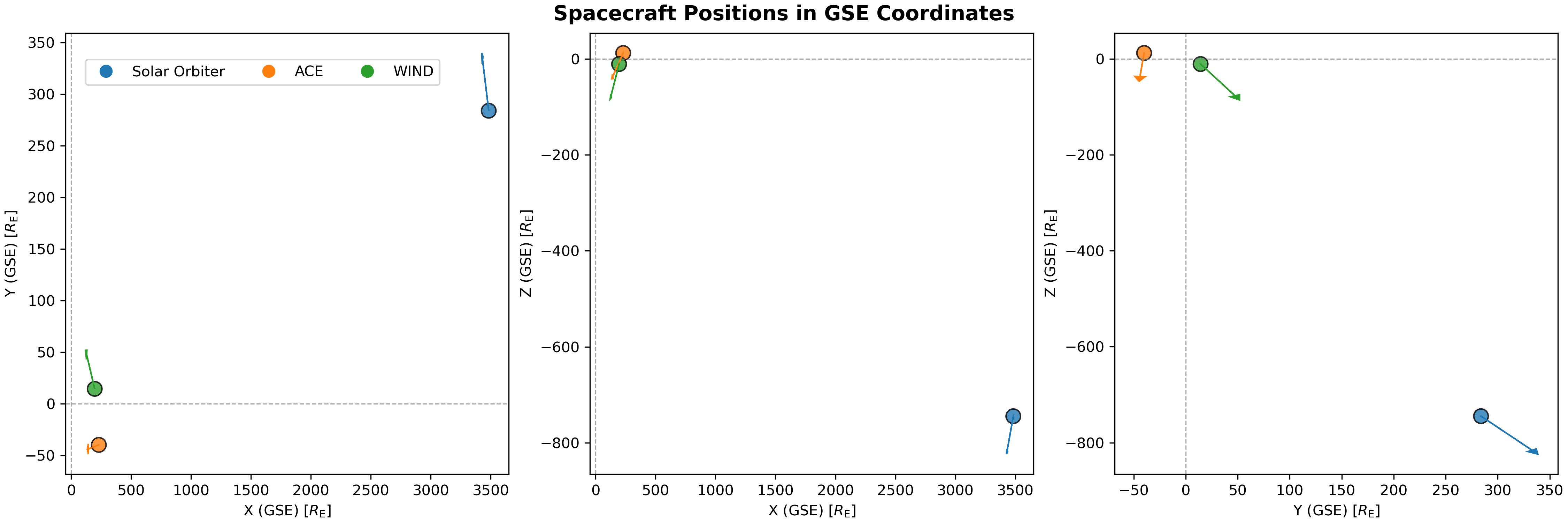}
   \caption{Orbital configuration for the event, showing the positions of Solar Orbiter (blue), ACE (orange) and \textit{Wind} in the GSE coordinate system when the IP shock was recorded at Solar Orbiter on 3 November 2021 14:04:26 UT. The arrows show the (average) shock normal vectors.}
              \label{fig:orbit}%
\end{figure*}

A fundamental quantity to address how particle acceleration operates at IP shocks are the intensity-time profiles of energetic particles at various energies, which are often found to increase up to 100 times in correspondence with the shock passage~\citep{Bryant1962}. These clear indications of efficient particle acceleration reveal not only the maximum energy achieved during the acceleration process, but also shed light on the mechanisms of the particle acceleration because they can be predicted by theoretical frameworks, for instance, by the diffusive shock acceleration theory~\citep[DSA,][]{Axford1977, Blandford1978, Bell1978a, Drury1983}.

A relatively small number of observations has been shown to agree with the predictions from the simplest form of DSA theory, however, such as that there should be an exponential rise in time-intensity profile of energetic particles up to the shock crossing, followed by a plateau-like profile, also called energetic storm particle (ESP) event~\citep[see][and references therein]{Ellison1985,Giacalone2012}. A plethora of particle responses is often found in association with IP shocks, including shock spike events\citep{Decker1983}, long-lasting upstream particle beams~\citep{Lario2022}, and flat upstream particle spectra~\citep{Perri2023}. Further, statistical studies performed using NASA Advanced Composition Explorer~\citep[ACE,][]{Smith1998} IP shocks observations elucidated that many IP shocks are not associated with any energetic particles response, while the majority of shocks show an irregular response that cannot be simply predicted~\citep{Lario2002}. These results indicate that the variability of the shock system and the medium through which it propagates are key for reconstructing the observational signatures of particle acceleration.

Shock space-time variability has been shown to be important at small scales (a fraction of up to some hundreds of ion skin depths) using multi-spacecraft observations, often complemented by simulations~\citep[e.g.,][]{Kajdic2019}. Spatial shock irregularities at larger scales are also very important, as shown by~\citet{Neugebauer2006}, where quasi-simultaneous ACE and~\textit{Wind}~\citep{Wilson2021} observations near-Earth revealed that energetic particle profiles differ for increasing spacecraft separations. These ACE--\textit{Wind} separations covered up to $\sim$3 Mkm (0.02 AU), which poses an upper limit to the scale of the variability that was possible to address.
The role of the shock evolution in the acceleration efficiency was studied early by \citet{Kallenrode1997}, who used the \textit{Helios} spacecraft~\citep{Schwenn1990} to sample IP shocks between 1 and 0.3 AU. Through a diffusion model, they revealed that energetic particle production exhibits variability based on the evolutionary stage of the shock, with some shocks being efficient accelerators of particles close to the Sun, while others showed the opposite behavior. This comprehensively shows that the shock space-time variability is far from being achieved, and other important effects such as large-scale turbulence that affect particle acceleration~\citep{Guo2021, Trotta2025a} are still elusive. This stimulates the need of multi-spacecraft observations with advanced instrumentation.

 \begin{figure*}  
   \includegraphics[width=0.99\textwidth]{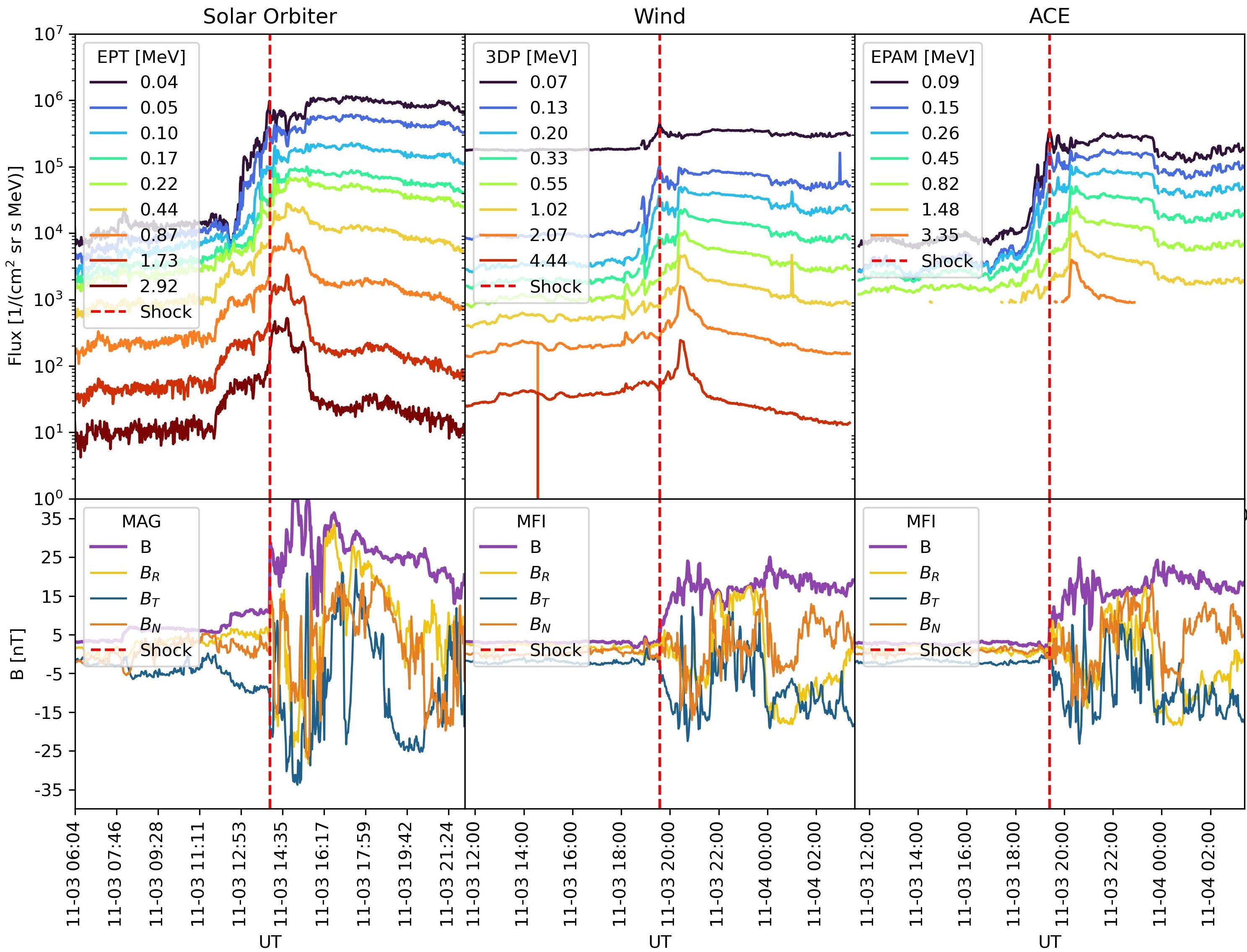}
   \caption{Sixteen-hour overview of energetic particle fluxes (top) and magnetic field (bottom) observations by Solar Orbiter, \textit{Wind}, and ACE (from left to right). The dashed vertical line identifies the shock transition, and it is at the center of the overview for each plot. We show 9 of the 64 energy channels available from Solar Orbiter EPT.}
              \label{fig:overview}%
\end{figure*}

A new spacecraft fleet, equipped with instruments yielding measurements at unprecedented time-energy resolution, is now available and investigates poorly explored regions in the heliosphere. In particular, the ESA Solar Orbiter~\citep{Muller2020} is investigating the inner heliosphere at distances of 0.3 -- 1 AU, thereby opening a new observational window for particle acceleration at IP shocks. In particular, the Energetic Particle Detector~\citep[EPD;][]{RodriguezPacheco2020} reveals novel detailed aspects of particle acceleration at IP shocks from thermal to high energies. A picture emerges according to which shock variability is a crucial component~\citep[e.g.,][]{Dimmock2023,Trotta2023c,Jebaraj2023b,Yang2024}. Further, the timely peak of the 11-year activity of solar cycle 25 provides us with a unique dataset of more than 100 IP shocks observed so far at different heliocentric distances, which complement and advance earlier classification efforts of the energetic particle response to IP shocks. A summary of these responses, as well as general IP shocks features and trends through the Solar Orbiter mission, is given by~\citet{Trotta2025b}.

Single-spacecraft observations of IP shocks are intrinsically limited. It is therefore important to note that Solar Orbiter joins a large existing fleet of spacecraft that orbits near Earth and in poorly investigated regions of the heliosphere. This effectively introduces a new golden era for direct spacecraft investigations. For example, joint observations of Solar Orbiter and the NASA Parker Solar Probe~\citep[PSP;][]{Fox2016} enable us to investigate previously inaccessible evolutionary features of solar eruptive phenomena~\citep{Trotta2024, Palmerio2025} and to study the more general question of the origin of the solar wind and its propagation~\citep[e.g.,][]{Horbury2023,Rivera2024,Wang2025}. The multiple vantage points enable us to exploit particular spacecraft configurations and to observe the same plasma as it propagates out in the heliosphere with separations that no single mission can achieve~\citep[see][]{Laker2024, Trotta2024b}.

We exploit a radial alignment between Solar Orbiter and the ACE and \textit{Wind} spacecraft near Earth, which observed a strong shock at 0.8 and 1 AU on the 3 November 2021. This event was previously studied with a focus on the small-scale plasma dynamics in the shock surroundings and produced the first multi-spacecraft observations of transient shocklets upstream of the IP shock~\citep{Trotta2023a}. It also represents the first observations of localised dynamic pressure enhancements, namely downstream jets, for IP shocks~\citep{Hietala2024}.\citet{Yang2023} focused on the production of suprathermal protons at Solar Orbiter, while ~\citet{Chiappetta2023} focused on the particle energy spectra and magnetic fluctuation transmission from upstream to downstream of the shock using Solar Orbiter and three other near-Earth spacecraft. 

We studied the response of particle fluxes at different energies to the IP shock crossing at Solar Orbiter and at the well radially aligned ACE and \textit{Wind}. We devised a novel technique to quantitavely address the variability in the observed intensity-time profiles, which we placed in the context of the shock-intrinsic space-time variability. We show that a) the effect of shock evolution makes the acceleration to the highest ($\geq 1$ MeV) energies less efficient as the shock propagates from 0.8 to 1 AU, with profiles that appear to be delayed with respect to the shock crossing at these energies, and that b) the spatial variability associated with the shock system affects the time at which energetic particles peak and the fine structure of the particle intensity-time profiles, in particular, for ions with energies $\lessapprox 0.5$ MeV.

\begin{figure}  
   \includegraphics[width=0.49\textwidth]{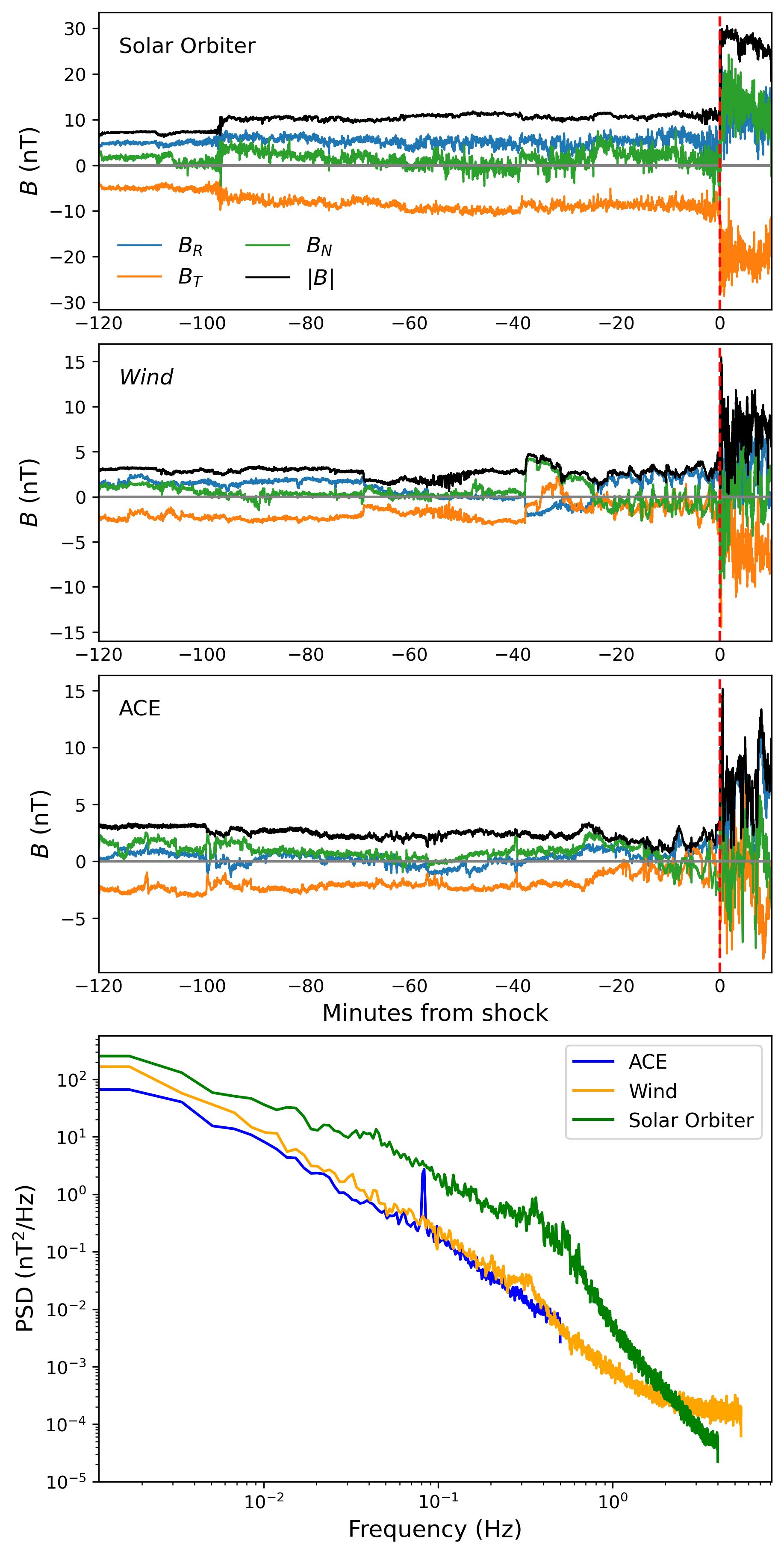}
   \caption{Magnetic field magnitude and components for 2 hours upstream of the shock at Solar Orbiter, \textit{Wind}, and ACE (top to bottom, respectively). The dashed red line marks the shock transition. In the bottom panel, we show the trace magnetic field power spectral densities for 2 hours upstream of each event. }
              \label{fig:foreshocks}%
\end{figure}

\begin{figure*}  
   \includegraphics[width=0.99\textwidth]{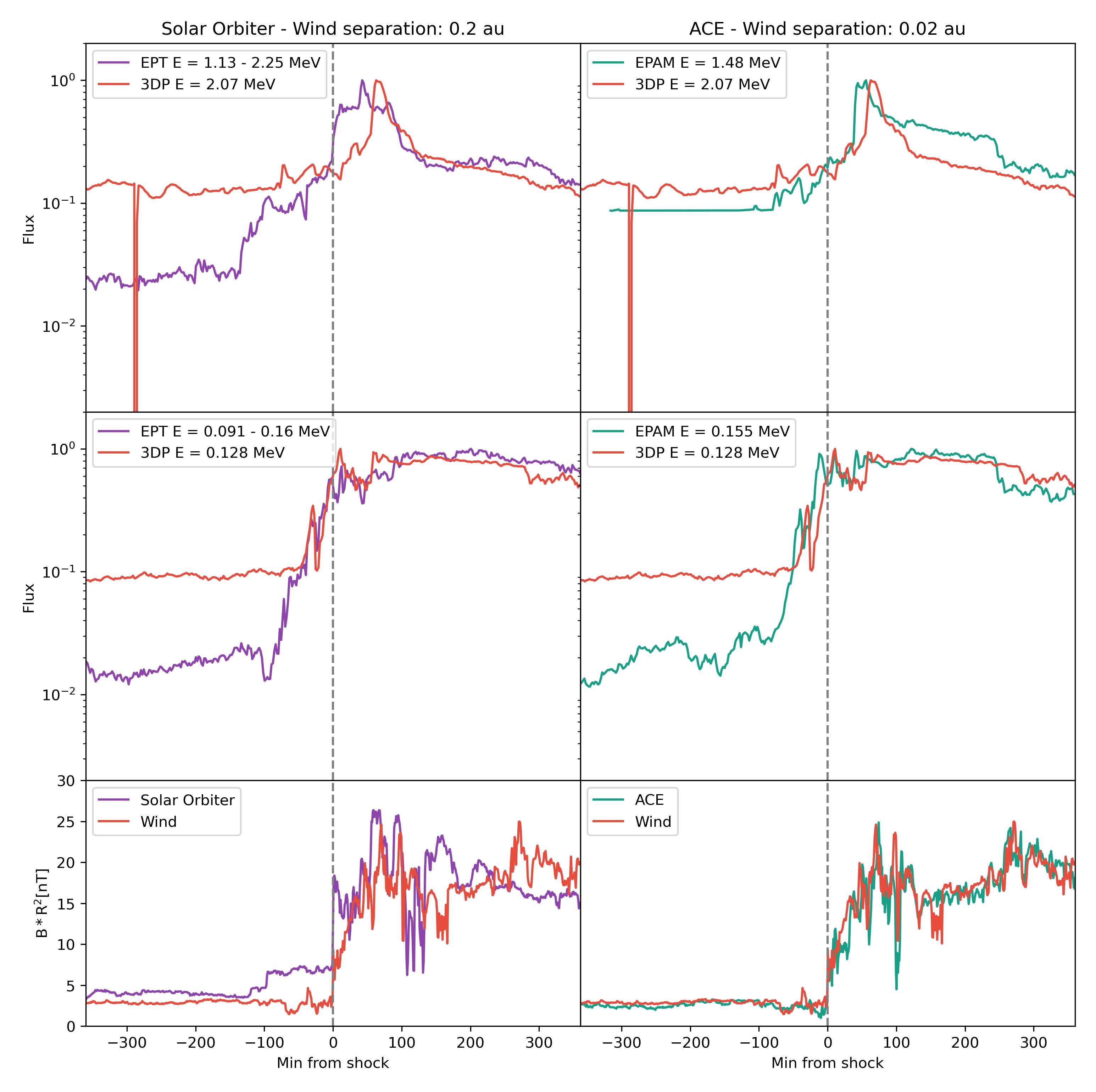}
   \caption{Twelve-hour overview of the event, where magnetic fields and intensity-time energetic particle profiles are overlaid for the Solar Orbiter-\textit{Wind} (left) and for the ACE-\textit{Wind} observations. The fluxes are normalized to the maximum flux in each interval. From top to bottom, we show an example of high (top) and low (middle) energy particle intensity-time profiles and the magnetic field magnitude, scaled with the square of the heliocentric distance.}
              \label{fig:overlay}%
\end{figure*}

The paper is organised as follows: In Section~\ref{sec:data} we present the datasets. In Section~\ref{sec:results} we present and discuss our results. Section~\ref{sec:conclusions} contains the conclusions.

\section{Data}\label{sec:data}

\begin{figure*}  
    \includegraphics[width=0.99\textwidth]{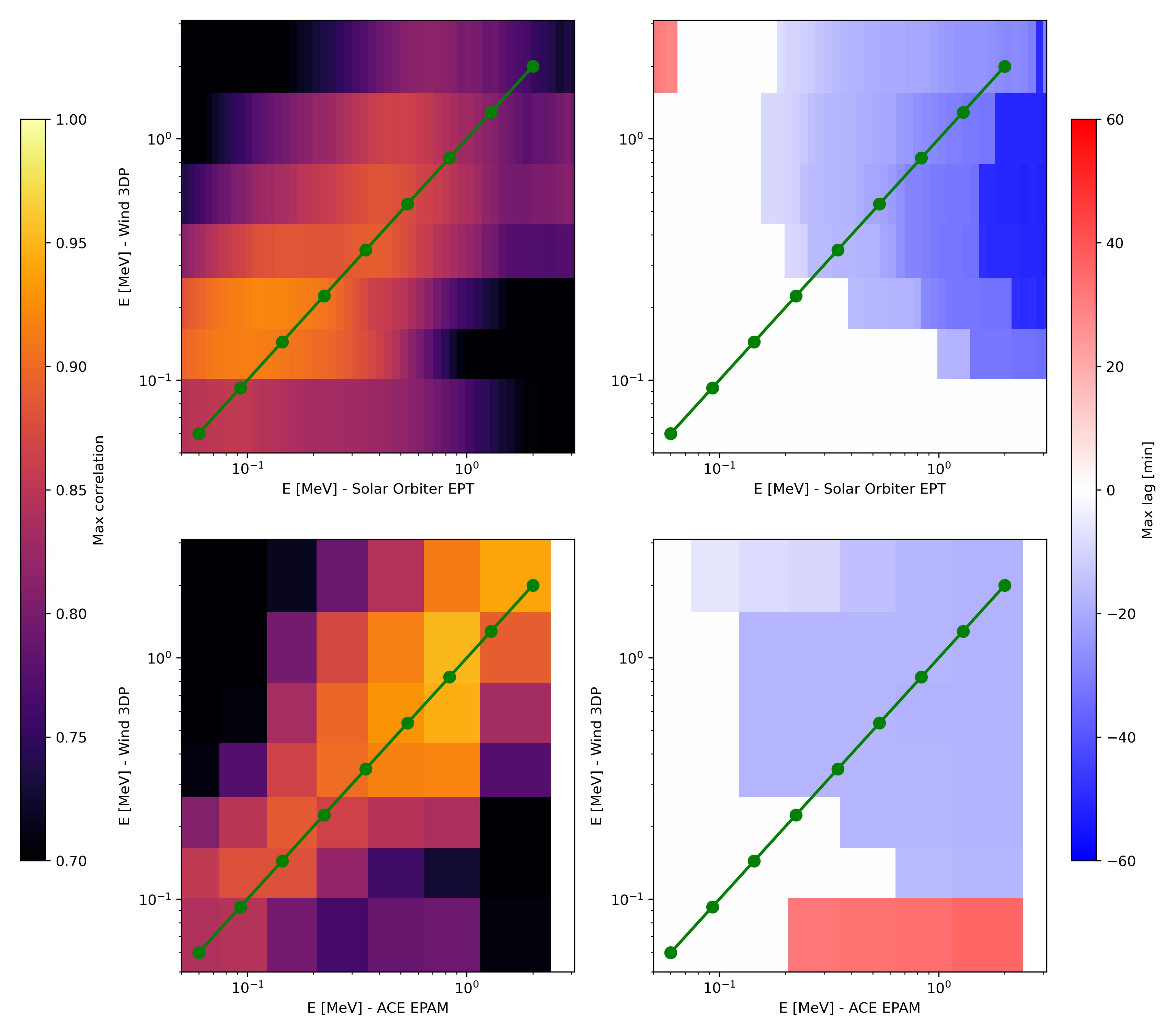}
    \caption{Left: Magic squares showing the peak correlation for particle intensity-time profiles in all available energy bins (left) for Solar Orbiter-\wind (top) and ACE-\wind (bottom). Right: Magic squares displaying the lag at which the peak correlation displayed in the left panels are achieved. The green dots show selected logarithmically spaced energies at which the peak correlation and lags are interpolated in Figure~\ref{fig:onedcorr} below.}
    \label{fig:magic}
\end{figure*}

In situ data from three different spacecraft were used in this work. For Solar Orbiter, we used magnetic field measurements
at a resolution of 64 vectors$/$s by the flux-gate magnetometer MAG \citep{Horbury2020}. Energetic proton fluxes were used 
as well, measured from the sunward aperture of the Electron and Proton Telescope (EPT-Sun) of the Energetic Particle Detector (EPD) suite~\citep[see][]{Wimmer2021} at a resolution of 1 s in 64 energy channels between 60 and 7000 keV. For \textit{Wind}, we used data from the  \wind 
Magnetic Field Investigation (MFI) at a resolution of 11 vectors$/$s \citep{Lepping1995}, and energetic ion fluxes from the 
Three-Dimensional Plasma Analyzer~\citep[][]{Lin1995}  at 24s resolution in eight energy channels from 70 to 6800 keV. From ACE, we 
used the magnetic field experiment  at a resolution of 1 s \citep{Smith1998} and of the Electron, Proton and Alpha Monitor~\citep[EPAM;][]{Gold1998} at a resolution of 12 s with seven channels in the energy range of 47 -- 4800 keV.

\section{Results and discussion}\label{sec:results}

On 3 November 2021 a strong fast-forward IP shock crossed Solar Orbiter at 14:04:26 UT. The event was later detected by several near-Earth spacecraft. Figure~\ref{fig:orbit} shows the orbital configuration of the event, where Solar Orbiter has a latitude-longitude separation to the Earth of 1.1-2.0 degrees. The event was detected by ACE and \textit{Wind}, which orbited the Lagrange point L1, at 19:24:05 and 19:35:01, respectively. The shock was in a supercritical regime at Solar Orbiter and near-Earth because the Alfv\'enic ($\rm{M_A}$) and fast magnetosonic ($\rm{M_{fms}}$) Mach numbers were estimated to be  $\rm{M_A} \sim 6.2, 5.6$ and $\rm{M_{fms}} \sim 5.5, 5.3$ at Solar Orbiter and \textit{Wind}, respectively. We note that this IP shock has higher Mach numbers than most IP shocks, which are typically in the subcritical regime with M close to unity~\citep[e.g.,][]{Kilpua2015_shocks,PerezAlanis2023,Trotta2025b, Kruparova2025}. The angle between the upstream magnetic field and the shock normal direction, $\theta_{Bn}$, which is very important for collisionless shock dynamics, was computed at all three locations, with the highest local value at Solar Orbiter ($\theta_{Bn} \sim 45^\circ$). At \textit{Wind} and ACE, the shock was closer to the so-called quasi-parallel regime, with $\theta_{Bn}\sim 33^\circ, 10^\circ$, respectively. These shock parameters were computed using single-spacecraft techniques (mixed mode for the shock normal, and mass flux conservation for the shock speed~\citep{Paschmann2000}) and employing a systematic variation in the upstream-downstream averaging windows according to the method outlined by~\citet{Trotta2022b}. A full characterization of the shock crossing at each spacecraft can be found in~\citet{Trotta2023a}. Figure~\ref{fig:orbit} shows that the locally computed shock normals (arrows) vary significantly spatially, which indicates that the shock front is, as expected, corrugated due to both self-induced and ambient fluctuations~\citep[][]{Kajdic2021JGR,Trotta2023b}.

\begin{figure*}  
    \includegraphics[width=0.99\textwidth]{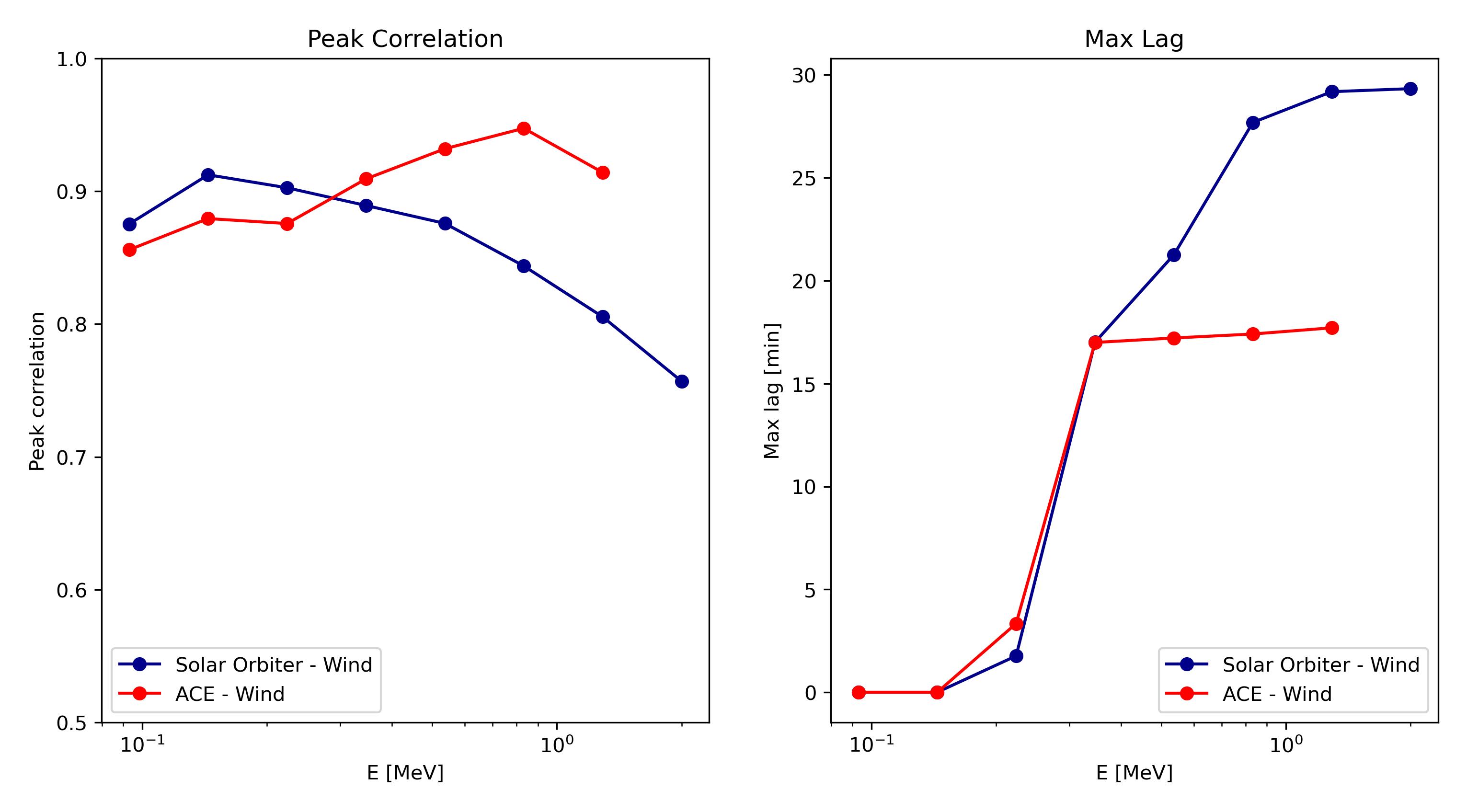}
    \caption{Peak correlation (left) and lag (right) as a function of energy, interpolated from the magic squares (see green dots in Figure~\ref{fig:magic}).}
    \label{fig:onedcorr}
\end{figure*}

Figure~\ref{fig:overview} shows an overview of the event, with energetic proton intensity-time profiles (top) and magnetic fields (bottom) as observed by Solar Orbiter, \textit{Wind}, and ACE. At all spacecraft, a typical ESP signature, namely a rise in the energetic particles fluxes, is observed before the shock transition, followed by a plateau after the shock, for energies up to about 0.5 MeV. Higher energies instead show a spike-like behavior. \citet{Chiappetta2023} studied the upstream and downstream energy spectra and concluded upon fitting with Band functions that lower-energy particles are accelerated in a DSA-like regime, while higher-energy particles are accelerated stochastically in the shock downstream close to the shock transition. As noted in previous studies as well~\citep{Yang2023}, a high level of variability is observed for the magnetic field both upstream and downstream of the shock.

To further characterize the magnetic environment of the shock upstream at each spacecraft, we show in Figure~\ref{fig:foreshocks} a zoom of the magnetic field magnitude and its components for the event at Solar Orbiter, \wind, and ACE (top to bottom). The upstream magnetic environment of the shock is different at the Solar Orbiter crossing than at 1 AU. At Solar Orbiter, an extended region of weakly compressive fluctuations can be seen upstream of the shock. For this crossing, we also note a previous weak shock at 12:28:30 about 90 minutes upstream of the shock.  $\rm{M_A}$ and $\rm{M_{fms}}$ are 1.9 and 1.8, respectively, the shock speed in the spacecraft rest frame is about 430 km/s, and the \tbn has been estimated to be 41$^\circ$ using MX3~\citep{Trotta2025a}. Albeit small, this event may play an important role in terms of precontitioning of the Solar Orbiter upstream magnetic environment characterized by high levels of fluctuations. The shock indeed caused a stationary increase in the solar wind magnetic field fluctuations reminiscent of the transition of a solar wind boundary. A weak shock like this might also play a role in preconditioning the suprathermal tails of the upstream particle distribution for the shocks at Solar Orbiter\citep[e.g.,][]{Nyberg2024}.

\wind and ACE observations reveal different features of the upstream magnetic environment, with a region of compressive fluctuations known as shocklets that is compatible with the fact that the shock \tbn is lower for the 1 AU crossings~\citep[see][for further details]{Trotta2023a}. In the bottom panel of Figure~\ref{fig:foreshocks}, we show the trace power spectral density for the magnetic field, collected in the two hours prior to the shock. These spectra characterize the environment through which the IP shock propagates, with a preexisting turbulent spectrum typical of the solar wind~\citep[see][]{BrunoCarboneReview}. The spectrum at Solar Orbiter is more intense by one order of magnitude than the two very similar near–Earth upstream spectra. For frequencies down to 0.01 Hz, the spectral slope is compatible with a Kolmogorov scaling, and it becomes flatter at lower frequencies. Embedded in these typical solar wind conditions, shock–induced foreshock fluctuations are visible as enhancements in the PSDs at .1-1 Hz frequencies~\citep[see][]{Guo2021}. This enhancement is more prominent at Solar Orbiter, possibly due to the presence of the weak upstream shock event. The fact that the whole PSD (and therefore the magnetic fluctuations) is at a higher level at Solar Orbiter is expected because of the radial evolution of solar wind magnetic fluctuations~\citep[see][]{Matteini2024}. This behavior has important implications for the energetic particle response to the shock passage, with stronger fluctuations providing a stronger source of scattering, thereby enhancing the possibility of energetic particles to interact with the shock for longer times.

We now focus on the variability of the energetic particle profiles and stress two particular aspects: First, Figure~\ref{fig:overview} shows that the largest increase in the energetic particle flux is observed at the shock location for Solar Orbiter (left), while for \text{Wind} and ACE, it takes place in the shock downstream. This is particularly evident for high-energy ($\gtrapprox 0.5$ MeV) particles. Second, even at Solar Orbiter, the peak in energetic particle flux is reached downstream of the shock (see the spike feature at around 15:30 UT at Solar Orbiter and 21:00 UT at \textit{Wind} and ACE). Finally, the observed profiles of the energetic particles are modulated by the magnetic field fluctuations both upstream and downstream, as is the case for the particle dropout that is particularly evident at low energies upstream.

Figure~\ref{fig:overlay} shows a direct comparison of particle intensity--time profiles and their variability. We show a 12-hour overview of the event centered at the time of the shock crossing, and we overplot energetic particle profiles in two different energy ranges, with high ($\sim$ 2 MeV) energies in the top panels and lower ($\sim$ 0.15 MeV) energies in the middle panels. The left panel shows the event at a separation of 0.2 AU for Solar Orbiter and \textit{Wind}, and the right panel shows it for ACE and \wind (0.02 AU separation). As we are interested in the variability in the energetic particle fluxes and not in their absolute levels, we normalized each intensity-time profile to its relative peak flux in the chosen time interval. Finally, in the bottom panel, we show the profile of the magnetic field magnitude, scaled with the square of heliocentric distance to take the general trend of the solar wind expansion into account~\citep[e.g.,][]{Maruca2023}. Thisanalysis revealed several interesting features that are not readily seen in the Figure~\ref{fig:overview} plot.

First, we note that magnetic field profiles at the ACE--\wind separations agree well, although some differences arise at small scales (see Figure~\ref{fig:foreshocks}). While the behavior of the energetic particles is not expected to be strongly affected by these irregularities, they are shown to be important for the thermal plasma dynamics, where they might generate out-of-equilibrium velocity distribution functions~\citep[e.g.,][]{Cohen2019, Preisser2020,Dimmock2023} and for the acceleration of particles into their suprathermal ($\sim$ 10 keV) regime, where it was shown that they may lead to an irregular injection process in both space and time\citep{Trotta2023c}. At larger separations (Figure~\ref{fig:overlay} left), stronger differences in the magnetic field profile arise, while a general trend is still recovered, especially in some portions of the shock downstream, including the close vicinity of the shock. 
Inspecting mildly energetic ($\sim$ 0.15 MeV) particles (Figure~\ref{fig:overlay}, middle), we find that their overall behavior across the shock transition is remarkably similar at .2 and .02 AU separations. This is an indication that particle acceleration operates similarly at all three vantage points. Differences at small scales are present as a result of the local conditions of the plasma, however. To assess these small-scale differences, a spacecraft constellation with smaller separations such as ESA \textit{Cluster}~\citep{Escoubet1997} is needed. This underlines again that multi-spacecraft observations are needed to understand the particle acceleration in the heliosphere.

For energetic particles ($\sim$ 2 MeV), we find a different behavior. The ESP feature of the profile becomes less evident and transitions into an irregular spike-like event. As noted above, the largest flux jump is at the shock ramp for Solar Orbiter, but occurs in the shock downstream for both \wind and ACE. We interpret this as an evolutionary effect of the shock, which has likely a decreased capability to accelerate particles to the highest energies at 1 AU. Thus, the high energy particles observed at 1 AU were likely accelerated at a faster rate earlier, when the shock was in the inner heliosphere, and the rate of acceleration slowed as the shock moved outward. As the high-energy particles created earlier scatter in pitch angle in the shocked plasma behind the shock, they are advect with the speed of the plasma, which is slower than the shock. Thus, their peak is behind it. In addition to this evolutionary effect, Figure~\ref{fig:overlay} highlights the behavior of the peak in energetic particles, which occurs in the shock downstream. This is common for IP shocks, as recently elucidated using a large statistical set from Solar Orbiter~\citep{Trotta2025b}. The figure shows that a peak drifts further downstream as the shock propagates from Solar Orbiter to \wind. This important effect is mediated by the magnetic variability in the shock downstream. We propose that the drifting spike is observed through the temporary connection of the spacecraft to a portion of the shock front that efficiently produces high-energy particles, in a similar fashion to what was suggested for single-spacecraft observations by \citet{Giacalone2021,Yang2025}. Thus, the shock and downstream structure appear to play a crucial role in explaining the observed intensity-time profiles of these energetic particles. A similar behavior was observed for different energy channels, but is not displayed in the figure for clarity.

Our discussion is limited by the fact that it remains based on somewhat qualitative interpretations of the observations. Together with the growing spacecraft fleet that can directly observe energetic particles in the heliosphere with a great level of detail and at separations that no single mission can achieve, this motivated us to develop a quantitative diagnostic addressing the variability in energetic particles.

The new diagnostic is presented in Figure~\ref{fig:magic}, where we study the (normalized) peak correlation between particle intensity-time profiles for all the available energy channels. In this way, it is possible to construct a magic square in which well-correlated intensity-time profiles are easily identified (Figure~\ref{fig:magic} left). A similar magic square is constructed with the time lag at which the peak correlation is achieved, which helps us to identify drifting peaks and other time-shifted features of the particle profiles.

Considerable information can be extracted from Figure\ref{fig:magic}. First, the overall correlation of energetic particle profiles is lower for Solar Orbiter and\wind at a separation of 0.2 AU (top left panel). Further, in this case, it is possible to observe that higher energies are less well correlated. This effect is related to the DSA-like plateau that transitions to a spike-like response from Solar Orbiter to \wind. The relative magic-square lag (top right panel) captures the drifting of the spike feature in the energetic particle profiles discussed qualitatively above (Figure~\ref{fig:overlay}). Conversely, the ACE-\wind magic square shows that energetic particle profiles are generally correlated well, with a slight increase at higher energies (Figure~\ref{fig:magic}, bottom left). This is due to the small-scale variability that affects the dynamics of lower-energy particles. For a 0.02 AU separation, a weak dependence of the lag on energy is found that captures the drift of the flux peak that is probably due to the spatial variability associated with the shock downstream. 

The diagnostic presented here enabled us to extract information from multi-mission datasets where a common problem derives from the fact that energy channels often do not overlap exactly~\citep{Dresing2023}. For instance, by comparing the top and bottom squares in Figure~\ref{fig:magic}, the revolutionary level of detail in Solar Orbiter observations is evident. It is then natural to reduce the information from the magic-square diagnostic to a simpler diagnostic. This is what we did in Figure~\ref{fig:onedcorr}, where we linearly interpolated the peak correlation and lags at nine logarithmically spaced energies compatible with the spacing in energy that \wind and ACE have (green dots in Figure~\ref{fig:magic}). The left panel of Figure~\ref{fig:onedcorr} shows that the particle decorrelated at high energies when the spacecraft separation was 0.2 AU (Solar Orbiter-\wind), and had a weaker and opposite energy dependence at smaller 0.02 AU separations (ACE-\wind). The right panel of Figure~\ref{fig:onedcorr} shows that at high energies, a larger drift is identified in the particle time-intensity profiles for larger separations.

\section{Conclusions}\label{sec:conclusions}

The energetic particle intensity-time profiles that are observed directly by means of spacecraft instrumentation contain invaluable information about the elusive mechanisms of particle acceleration in the heliosphere. One the one hand, previous studies elucidated the importance of studying the evolutionary features of these energetic particle profiles in response to the passage of IP shocks, but they were limited by instrumental capabilities, and focused so far on the statistical treatment of events at different heliocentric distances~\citep{Kallenrode1997}. On the other hand, previous studies also showed using single- and multi-spacecraft data near Earth that the spatial structure of IP shocks is fundamental to explain how particles are energized in space and in astrophysical systems~\citep{NeugebauerGiacalone2005, Neugebauer2006}.

We focused on the space-time variability of energetic ion intensity-time profiles by exploiting a three-spacecraft configuration with separations that no single mission can resolve. Our target was a strong IP shock observed near Earth by \wind and ACE and at 0.8 AU by the well radially aligned Solar Orbiter. This event constitutes an important dataset of Solar Cycle 25 and was the target of other studies with different foci~\citep{Trotta2023a,Yang2023,Chiappetta2023,Hietala2024}.

We focused on the 0.01 - 5 MeV energy range and found that particles with a different energy have different dynamics upon their interaction with the shock. At energies $\lessapprox$ 0.5 MeV, the energetic particles exhibited an ESP-like response that was compatible with diffusive acceleration at the shock front, while higher energies tended to exhibit a spike-like profile. A crucial feature of this observations is that the largest rise in energetic particle fluxes occurred at the same time as the shock crossed Solar Orbiter, and downstream with a delay of about 30 minutes for the near-Earth spacecraft. We suggest that this is due to an evolutionary effect of the IP shock that is unable to efficiently accelerate particles to the highest energies as it propagate from 0.8 to 1 AU. This scenario is compatible with previous studies that combined in situ observations and numerical modeling and performed statistical studies of the radial dependence of the particles~\citep{Lario1998, Lario2006}, which we now tested for the same shock. 

The direct comparison of the same event enabled us to characterize the role of the magnetic fluctuation environment in shaping energetic particle responses, which is an important ingredient to discuss for these observations. At all spacecraft, the upstream magnetic environment was found to be a combination of preexisting and shock-induced fluctuations. These fluctuations are the scattering sources for particles that participate in the process of acceleration to high energies. Thus, it is  important to note that low-energy particles are more effectively confined in the shock surroundings for longer times than the high-energy particles, which help us understand the observations that low-energy particle fluxes have their largest jump at the shock location in all crossings. From this point of view, we note that energetic particles at Solar Orbiter interact with both a younger shock and with a higher level of fluctuations due to the evolved solar wind, which may explain why particles with high energies peak at the shock at Solar Orbiter and downstream in the other observations. We showed that the cross-correlation of energetic particle intensity-time profiles is a powerful diagnostic to quantitatively address the behavior discussed above (see Figures~\ref{fig:magic} and \ref{fig:onedcorr}).

We used the same diagnostic to address how the  spike-like feature found at high ($\gtrapprox$ 0.5 MeV) energies at all spacecraft, modulated by strong plasma structuring in the shock downstream, drifts further downstream at separations of 0.2, and it was found at different times with respect to the shock when comparing \wind and ACE observations as well. We conclude that this behavior is a consequence of the intrinsic space-time variability associated with the shock system, where a transient connection to different portions of the shock front that accelerates particles to the highest energies is often observed~\citep[e.g.,][]{Yang2025}.  We also note that particle trapping in magnetic field structures such as flux ropes may play an important role for the further acceleration of particles that propagate downstream of the shock~\citep[e.g.,][]{Zank2015,Kilpua2023,Pezzi2024}.

A key finding of this work is that in order to understand how the shock acceleration of particles operates at different evolutionary stages, not only the shock must be considered as evolving, but the environment through which it propagates as well. We then elucidated that large-scale irregularities shape the energetic particles profile by resolving a range of separations that cannot be tackled with a single mission, and this was previously done only statistically for different events~\citep[see][]{Kallenrode1997}.

As we are now in a new golden era for multi-spacecraft studies in the heliosphere and target a quantitative characterization of energetic particle variability at IP shocks, it is critical to develop a new technique for cross-correlating energetic particle profiles and extracting information about the changes in particle production for the shock system at different temporal and spatial locations to quantitavely characterize particle acceleration and its variability. This technique highlighted that the particle profiles are well correlated at low energies and are decorrelated at higher energies ($\gtrapprox$ 0.5 MeV) at separations of 0.2 AU, and have a weak and opposite dependence for a separation of 0.02 AU. The added value of this technique is that it overcomes instrumental limitations from multi-mission studies in which the energy channels do not overlap exactly. 

From this point of view, this study is the staring point of further multi-spacecraft investigations of particle acceleration at IP shocks, which will involve more events and possibly larger separations in the future (e.g., using lineups of Earth--Solar Orbiter--PSP). Further, this study paves the way for the application for the growing near-Earth spacecraft fleet that was recently joined by the ISRO \textit{Aditya-L1} spacecraft~\citep{Seetha2017} and is soon to be joined by the NOAA Space Weather Follow On Lagrange 1 (SWFO-L1) mission and the NASA Interstellar Mapping and Acceleration Probe~\citep[IMAP;][]{McComas2018}.

\begin{acknowledgements}
    DT acknowledges fruitful discussions with D. Lario, A. Szabo, D. Turner and P. C. Escoubet. JG acknowledges support from NASA under grants 80HQTR21T0005 and 80NSSC20K1815, and from the IS$\odot$IS instrument suite on NASA's Parker Solar Probe Mission under contract NNN06AA01C, and the NSF under grant 1931252. DT would also like to welcome Enrico Trotta Verrina to planet Earth.

\end{acknowledgements}

\bibliographystyle{aa}
\bibliography{bibby}

\end{document}